\documentclass[%
 reprint,
   aps,
  prfluids,
  onecolumn,
  amsmath,
  amssymb,
]{revtex4-2}
\usepackage[english]{babel}
\usepackage{graphicx}
\usepackage{dcolumn}
\usepackage{bm}
\usepackage{hyperref}
\usepackage{cleveref}
\usepackage[caption=false]{subfig}
\usepackage{xkeyval,xcolor}
\usepackage{subfig}
\usepackage{pdfcomment}
\crefname{section}{sec.}{secs.}
\crefname{appendix}{appx.}{appx.}

\makeatletter
\newlength{\sfp@hseplen}\newlength{\sfp@vseplen}
\define@cmdkey{subfigpos}[sfp@]{pos}[ul]{}
\define@cmdkey{subfigpos}[sfp@]{font}[\small]{}
\define@cmdkey{subfigpos}[sfp@]{vsep}[2\baselineskip]{\setlength{\sfp@vseplen}{\sfp@vsep}}
\define@cmdkey{subfigpos}[sfp@]{hsep}[10pt]{\setlength{\sfp@hseplen}{\sfp@hsep}}
\newcommand{\subfigimg}[3][,]{%
  \setkeys{Gin,subfigpos}{pos,font,vsep,hsep,#1}
  \setbox1=\hbox{\includegraphics{#3}}
  \ifnum\pdfstrcmp{\sfp@pos}{ul}=0
    \leavevmode\rlap{\usebox1}
    \rlap{\hspace*{\sfp@hsep}\raisebox{\dimexpr\ht1-\sfp@vsep}{\sfp@font{#2}}}
    \phantom{\usebox1}
  \else\ifnum\pdfstrcmp{\sfp@pos}{ur}=0
    \leavevmode\usebox1
    \llap{\raisebox{\dimexpr\ht1-\sfp@vsep}{\sfp@font{#2}}\hspace*{\sfp@hsep}}
  \else\ifnum\pdfstrcmp{\sfp@pos}{lr}=0
    \leavevmode\usebox1
    \llap{\raisebox{\sfp@vsep}{\sfp@font{#2}}\hspace*{\sfp@hsep}}
  \else
    \leavevmode\rlap{\usebox1}
    \rlap{\hspace*{\sfp@hseplen}\raisebox{\sfp@vsep}{\sfp@font{#2}}}
    \phantom{\usebox1}
  \fi\fi\fi
}
\makeatother

\newcommand{\bracket}[1]{\langle {#1} \rangle}

\newcommand{\di}[0]{\,\textrm{d}}

\newcommand{\reynolds}[0]{\mathrm{Re}}
\newcommand{\nusselt}{\mathrm{Nu}}
\newcommand{\rayleigh}{\mathrm{Ra}}
\newcommand{\prandtl}[0]{\mathrm{Pr}}

\begin{document}

\preprint{APS/123-QED}

\title{Transition to the ultimate regime of turbulent convection in stratified inclined duct flow}

\author{Rundong Zhou$^{1}$}
\author{Adrien Lefauve$^{2,3}$}
\author{Roberto Verzicco$^{1,4,5}$}
\author{Detlef Lohse$^{1,6}$}
\affiliation{%
 \footnotesize{$^1$Physics of Fluids Department, J.M. Burgers Center for Fluid Dynamics and Max Planck Center Twente for Complex Fluid Dynamics, Faculty of Science and Technology, University of Twente, P.O. Box 217, 7500 AE Enschede, Netherlands \\
$^2$Grantham Institute -- Climate Change and the Environment, Imperial College, London SW7 2AZ, UK \\
$^3$Department of Civil and Environmental Engineering, Imperial College, London SW7 2BU, UK \\
 $^4$Dipartimento di Ingegneria Industriale, University of Rome ‘Tor Vergata’, 00133 Rome, Italy \\
 $^5$Gran Sasso Science Institute, Viale F. Crispi, 7, 67100 L’Aquila, Italy\\
 $^6$Max Planck Institute for Dynamics and Self-Organization, Am Fassberg 17, 37077 Göttingen, Germany
}}%

\date{\today}

\begin{abstract}

The stratified inclined duct (SID) provides a canonical setup for sustained, buoyancy-driven exchange flow between two reservoirs of different density, and emerges as a new paradigm in geophysical fluid dynamics.
Yet, the flow dynamics remain unclear in the highly turbulent regime; laboratory experiments can access this regime but they lack resolution, while direct numerical simulations (DNS) at realistically high Prandtl number Pr\,=\,7 (for heat in water) have not achieved sufficiently high Reynolds numbers $\reynolds$. 
We conduct three-dimensional DNS up to $\reynolds= 8000$ and observe the transition to the so-called  ultimate regime of turbulent convection as evidenced by the Nusselt number scaling $\nusselt \sim \rayleigh^{1/2}$, indicating substantially enhanced transport. 
At the transition the shear Reynolds number, a key parameter characterizing boundary layer (BL) dynamics, exceeds the threshold range of 420 for turbulent kinetic BLs with the emergence of logarithmic velocity profiles. 
The nature of the transition towards ultimate SID flow is non-normal-nonlinear, i.e., subcritical and hysteretic, as is typical for the transition to fully turbulent shear flows.
Our work connects SID flow with the broader class of wall-bounded turbulent convection flows and gives insight into mixing properties in the vigorously turbulent
 regime encountered in oceanographic and industrial flows.
\end{abstract}

\maketitle

\section{Introduction}
Buoyancy-driven exchange flows occur in many natural and built environments when two large bodies of fluid at different densities are connected through a narrow channel. 
Such flows cause significant exchange of scalars (e.g., heat, salt, pollutants, nutrients) between the two reservoirs with minimal to no net volume transport~\cite{wood_lock_1970,wilkinson_buoyancy_1986}. 
They underpin key mixing and transport processes in oceanography and industrial flows. 
For example, the Mediterranean Sea depends on exchanges through the straits of Gibraltar and the Bosphorus~\cite{deacon_scientists_1971},  coastal environments depend on estuarine transport processes~\cite{fischer_mixing_1976,maccready_estuarine_2018}, and natural air ventilation in buildings and industrial safety scenarios depend on exchange between stratified air masses~\cite{linden_fluid_1999}. 
Moreover, these flows are inherently stably stratified shear flows, ideal for studying shear instabilities and sustained stratified turbulence, a research lineage extending back to seminal works by Osborne Reynolds~\cite{reynolds_experimental_1883} and G. I. Taylor~\cite{taylor_effect_1931}.

The stratified inclined duct (SID) experiment was developed to study such buoyancy-driven exchange flows with sustained shear under well-controlled conditions~\cite{meyer_stratified_2014}, serving as a continuous realization of Thorpe’s seminal closed tilted-channel experiments~\cite{thorpe_method_1968}. SID is schematically illustrated in  \cref{fig:SID}: 
\begin{figure}
\includegraphics[height=2.3in]{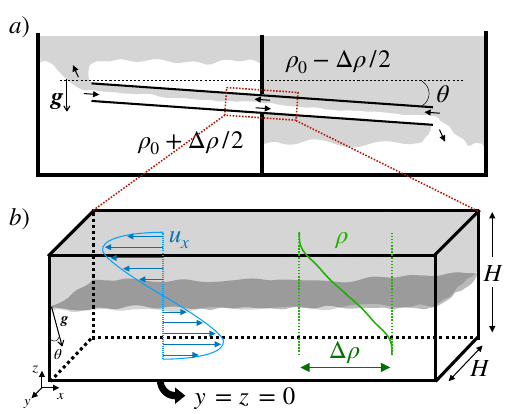}
\caption{\label{fig:SID} (a) Schematic illustration of the SID experiment.  (b) Detail of the mid-duct region in the duct reference frame, showing typical vertical profiles of mean streamwise velocity (blue) and density (green).}
\end{figure}
two reservoirs are filled with liquids of different densities $\rho_0 \pm \Delta\rho/2$, connected via a long rectangular duct tilted at an angle $\theta$ from the horizontal in the laboratory frame.
Once the duct is opened, a transient gravity current occurs, after which a sustained, buoyancy-driven, two-layer stratified exchange flow develops.
SID flow has emerged as a novel paradigm in fluid mechanics for exploring the routes to stratified turbulence, exhibiting rich transitional and intermittent dynamics which have been well documented over the last decade~\cite{meyer_stratified_2014,lefauve_regime_2019,lefauve_buoyancy-driven_2020,lefauve_experimental_2022,lefauve_experimental_2022-1,zhu_stratified_2023, lefauve_data-driven_2024, lefauve_routes_2025} (for a review, see \cite{lefauve_geophysical_2024}). 
With increased driving, either through larger density differences $\Delta\rho/\rho_0$ or greater inclination $\theta$, the flow typically transitions through four qualitatively distinct regimes---laminar, wave, intermittently turbulent, and fully turbulent.

\begin{figure*}[ht]
    \includegraphics[width=\textwidth]{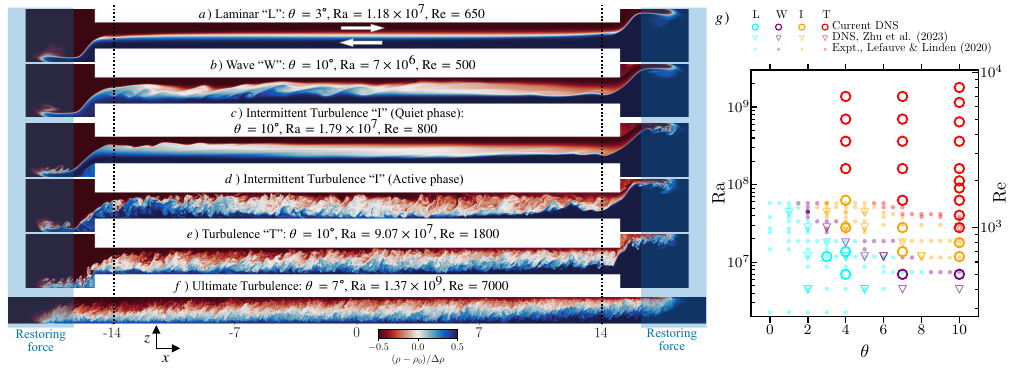}
    \caption{\label{fig:flow} (a-f)  Snapshots of the spanwise mid-plane density field $\rho$ for increasingly turbulent flow regimes. The flow directions are indicated by arrows in (a). The $z$-axis is scaled up by 1.5. Simulations include artificial restoring forces applied in the blue regions, see~\cref{sec:appendix}. The quantities are averaged within $x/H\in[-14,14]$ (black dashed lines) to eliminate edge effects. (g) Qualitative flow regime phase diagram illustrating the current DNS results alongside previous experimental~\cite{lefauve_buoyancy-driven_2020} and numerical~\cite{zhu_stratified_2023} data at $\prandtl=7$.}
\end{figure*}

The dimensionless Rayleigh number
\begin{equation}
    \rayleigh \equiv \frac{g(\Delta\rho/\rho_0)H^3}{\nu\kappa}
\end{equation}
quantifies the strength of the buoyancy driving, where $H$ is the duct height and serves as the characteristic length scale, $\nu$ the kinematic viscosity, $\kappa$ the thermal diffusivity, and $g$ the gravitational acceleration. 
Ra serves as a primary control parameter along with the inclination $\theta$. 
The hydraulic Reynolds number, sometimes used in the literature,
\begin{equation}
\reynolds \equiv \frac{\sqrt{g(\Delta\rho/\rho_0)H}(H/2)}{\nu}  = \frac{1}{2}\, \rayleigh^{1/2}\, \prandtl^{-1/2}
\end{equation}
is an equivalent parameter to $\rayleigh$ in this work as we fix the Prandtl number, $\prandtl \equiv \nu/\kappa = 7$, approximating heat in water around $20^\circ\mathrm{C}$. 
For salt in water, the corresponding $\prandtl_s\equiv\nu/D=700$ (also called Schmidt number Sc), with $D$ being the molecular diffusivity, is two orders of magnitude larger.
The velocity scale is the free-fall velocity $U_0 \equiv \sqrt{g(\Delta\rho/\rho_0)H}$, the typical observed peak velocity, while the lengthscale is $H/2$ in Re~\cite{lefauve_geophysical_2024}. The regimes are visualized by the density field snapshots and classified by the phase diagram in \cref{fig:flow}.

A key feature of SID flow is that its shear is generated exclusively by buoyancy, distinguishing it from canonical sheared convective flows that are typically driven mechanically~\cite{blass_flow_2020,zhou_diapycnal_2017,zhou_self-similar_2017}, or by external forcing mechanisms and imposed pressure gradients~\cite{yerragolam_scaling_2024,howland_turbulent_2024}. 
Additionally, it is bounded by no-slip sidewalls. 
These characteristics connect SID flow to other extensively studied canonical buoyancy-driven wall-bounded systems such as Rayleigh–B\'{e}nard (RB) convection (flow heated from below and cooled from above in an enclosed domain)~\cite{ahlers_heat_2009,lohse_small-scale_2010,chilla_new_2012,xia_current_2013,lohse_ultimate_2024,visakh_convection_2025}, axially homogeneous RB convection, also known as convection in a vertical channel (CVC)~\cite{gibert_high-rayleigh-number_2006,schmidt_axially_2012,pawar_two_2016,castaing_turbulent_2017},
vertical convection (flow between two vertical heated and cooled plates)~\cite{ng_vertical_2015,ng_changes_2017,ng_bulk_2018,wang_regime_2021,ke_turbulence_2023},
and Taylor-Couette flow (flow between two coaxial co- or counter-rotating cylinders)~\cite{dubrulle_momentum_2002,eckhardt_torque_2007,grossmann_highreynolds_2016}. 
In particular, for an extreme $\theta=90^\circ$, SID is a realization of CVC, where the heated/cooled plates of RB are replaced by reservoirs. For small $\theta$, SID retains distinct features of a stably stratified, strongly sheared mean exchange flow.
In spite of the recent interest, SID flow characteristics in the strongly turbulent regime have remained out of reach. Experiments at have reached $\reynolds = \mathcal{O}(10^4)$ at $\prandtl_s=700$ ~\cite{lefauve_buoyancy-driven_2020,lefauve_data-driven_2024}  but measurements of the velocity and density fields have so far been restricted to $\reynolds = \mathcal{O}(10^3)$ \cite{lefauve_experimental_2022,lefauve_experimental_2022-1}. Likewise, numerical simulations resolve the full flow fields but have also been restricted to $\reynolds= \mathcal{O}(10^3)$ at $\prandtl=7$~\cite{zhu_stratified_2023}.

In this work, we overcome these limitations by directly simulating highly turbulent SID flow up to $\reynolds=8000$, corresponding to $\rayleigh=1.8\times 10^9$, see \cref{fig:flow}g.
Our main finding is a pronounced enhancement in mass transfer efficiency beyond a threshold of $\rayleigh\approx 10^8$, with the scaling shifting from  $\nusselt\sim \rayleigh^{1/3}$ to $\nusselt\sim \rayleigh^{1/2}$, consistent with the so-called ultimate regime of turbulent thermal convection first proposed by Kraichnan~\cite{kraichnan_turbulent_1962} and found in homogeneous RB~\cite{schmidt_axially_2012}.
This transition is concomitant with the development of turbulent boundary layers (BL), suggesting a novel turbulent channel configuration.
These results provide the first observation of the ultimate regime in SID, and 3D numerical evidence of turbulent BLs in a purely buoyancy-driven system as previously found in vertical convection~\cite{ke_turbulence_2023}.

The the paper is organized as follows. 
In~\cref{sec:method}, we introduce the governing equations and the numerical schemes used to solve them. 
In~\cref{sec:results}, we present the main results on global response parameters: \Cref{sec:flux} demonstrates the transition to ultimate scaling in the mass flux; \cref{sec:mixing} quantifies the energy budget and mixing, and discusses quantities of particular interest to the oceanography and stratified turbulence communities; and \cref{sec:CVC} explains the physical origin of the observed ultimate scaling. 
\Cref{sec:BL} is on the local properties of the flow: We provide evidence for turbulent boundary layers in \cref{sec:reynolds,sec:profiles} and the non-normal-nonlinear nature of the transition in \cref{sec:transition}. 
The paper concludes in \cref{sec:conclusion}.

\section{\label{sec:method} Governing equations and numerical methods }
We numerically solve the imcompressible Navier-Stokes equations under Boussinesq approximation ($\Delta\rho/\rho_0\ll 1$). The flow is incompressible $\bm{\nabla}\cdot \bm{u} = 0$, and momentum and density transport equations are
\begin{subequations}
\label{eqn:NS}
    \begin{align}
        \label{eqn:momentum}\partial_t \bm{u} + (\bm{u}\cdot \bm{\nabla})\bm{u} &= - \bm{\nabla} p /\rho_0 + \nu \bm{\nabla}^2\bm{u} + (\rho-\rho_0) g\bm{\hat{e}}_\theta/ \rho_0 \ \\
        \label{eqn:scalar}\partial_t \rho + (\bm{u}\cdot \bm{\nabla})\rho &= \kappa \bm{\nabla}^2 \rho,
    \end{align}
\end{subequations}
where $\bm{u}=(u_x,u_y,u_z)$ is the velocity field. 
The coordinates are defined such that $x$-axis is aligned streamwise along the duct, $y$-axis spanwise across the duct, and $z$-axis normal to the duct, making an angle $\theta$ with the gravitational vector $\bm{g} = g \bm{\hat{e}}_\theta=g(\sin\theta,0,-\cos\theta)$. 
The duct streamwise aspect ratio (length/height) is 30 and the spanwise aspect ratio (width/height) is 1, as in most previous experimental and numerical studies. 
The boundary conditions are no-slip for velocity and no-flux for density at the four duct walls in $y$ and $z$.

The PDEs are solved with AFiD, a highly parallel solver with a second-order advanced finite difference scheme for spatial derivatives and a third-order Runge-Kutta scheme for time integration~\cite{verzicco_finite-difference_1996,poel_pencil_2015}. 
We follow the methodology introduced in~\cite{zhu_stratified_2023}, using immersed boundary methods (IBM) to replicate the duct-reservoir geometry (\cref{fig:flow}a-e) and artificial forces to allow for arbitrarily long runtime with modest domain size.
For simulations at higher $\rayleigh$, we remove the IBM boundary (\cref{fig:flow}f) to use a wall-clustered mesh in the $z$-direction to fully resolve the turbulent kinetic BLs~\cite{ceci_natural_2023}. 
Additionally, we use a multiple-resolution technique  to resolve $\rho$ for $\prandtl=7$ on a refined uniform mesh via a four-point Hermite interpolation method~\cite{ostilla-monico_multiple-resolution_2015}. 
With the same duct geometry and Pr as in the previous experimental~\cite{lefauve_buoyancy-driven_2020} and numerical studies~\cite{zhu_stratified_2023}, our simulations successfully reproduce the four qualitative flow regimes at the same control parameters $\rayleigh$ and $\theta$, see \cref{fig:flow}g.
In the overlapping region of parameter space, our results align closely with previous studies.
For further details of the DNS, we refer to \cref{sec:appendix}.

\section{\label{sec:results}Global transport and flow properties}
\subsection{\label{sec:flux}Mass flux}
\begin{figure*}
    \centering
    \subfloat{\includegraphics[height=2.1in]{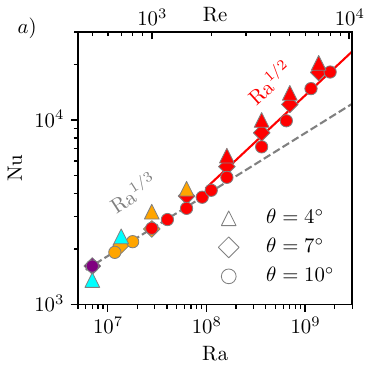}\label{fig:nux}}
    \subfloat{\includegraphics[height=2.1in]{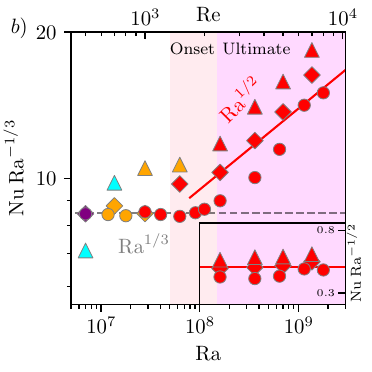}\label{fig:nux_comp}}
    \subfloat{\includegraphics[height=2.1in]{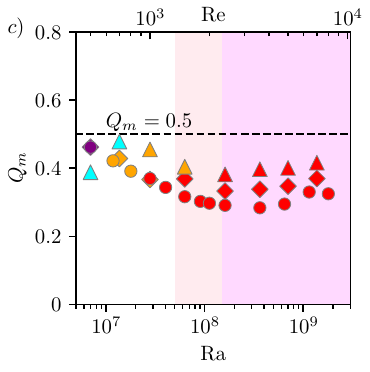}\label{fig:qm}}
    \caption{(a) Log-log plot of $\nusselt$ against $\rayleigh$. The fill color represents the flow regime as indicated in \cref{fig:flow}g and the shape corresponds to the inclination. The slope for small $\rayleigh$ is consistent with  $\nusselt\sim \rayleigh^{1/3}$ (dashed line). Scaling in the ultimate regime is guided by the red line $\rayleigh^{1/2}$. (b) Same data as (a), but compensated by $\rayleigh^{1/3}$. The inset shows the data in the ultimate regime compensated by $\rayleigh^{1/2}$. (c) Hydraulic dimensionless mass flux $Q_m$ against $\rayleigh$, the vertical scale is linear. The dashed line marks the hydraulic limit $Q_m= 0.5$.}
\end{figure*}
We begin by examining the primary observable of interest, the scalar transfer efficiency between the two reservoirs, quantified by the Nusselt number (the ratio of total transfer to purely conductive transfer;  also called
 Sherwood number when referring to mass transfer):
\begin{equation}
    \nusselt \equiv \frac{\bracket{u_x(\rho-\rho_0)}}{\kappa \Delta\rho /H} + 1,
\end{equation}
where $\langle\cdot\rangle$ denotes the volume- and time-average over the duct region of volume $V=28H^3$ between the black dashed lines in \cref{fig:flow}(a-f). 
In \cref{fig:nux}, we present $\nusselt$ as function of $\rayleigh$ in the range $\rayleigh\in[6\times10^6,2\times10^9]$ for three  angles $\theta=\{4^\circ, 7^\circ, 10^\circ\}$. Our results reveal two scaling regimes. 
For $\rayleigh\lesssim 10^8$, the data are consistent with a scaling of  $\nusselt\sim\rayleigh^{1/3}$, as expected for weakly driven thermal convection~\cite{lohse_ultimate_2024}. 
Beyond $\mathrm{Ra}\approx10^8$, the system undergoes a transition to the ultimate regime where the mass transfer efficiency follows the enhanced scaling $\rayleigh^{1/2}$. 

The transition is more apparent in \cref{fig:nux_comp}, where Nu is compensated by $\mathrm{Ra}^{1/3}$.  
In the low-Ra regime, data points scatter around the horizontal grey dashed line corresponding to $\nusselt\sim \mathrm{Ra}^{1/3}$, close to the effective scaling $\nusselt\sim\mathrm{Ra}^{0.3}$ reported in~\cite{pawar_two_2016} for weakly turbulent CVC.
Here, turbulence is not fully developed (or transitional) and SID flow is affected by a family of linear instabilities as reported in~\cite{atoufi_stratified_2023,zhu_long-wave_2024}, i.e., the first two points of $\theta=4^\circ$ deviate noticeably because the bulk flow structures are still laminar, as indicated by the fill color and flow visualization in \cref{fig:flow}.
The $\theta=10^\circ$ data follow the $\nusselt \sim \mathrm{Ra}^{1/3}$ scaling most closely, likely because the larger angle more closely resembles RB and CVC.
Castaing et al.~\cite{pawar_two_2016,castaing_turbulent_2017} attributed this regime to the absence of Kolmogorov inertial range, yet scalar turbulence is already developed for $\prandtl > 1$. 
Near the onset (highlighted by orange shading in \cref{fig:nux_comp}), the scaling departs from $\mathrm{Ra}^{1/3}$, 
and in the ultimate regime (magenta shading), data align with the red line $\mathrm{Nu}\sim\mathrm{Ra}^{1/2}$ for all three angles, highlighted in the inset. 
The ultimate regime reflects that the transport is independent of $\nu$ and $\kappa$. 
Under this assumption, the scaling is obtained by simple dimensional analysis \cite{kraichnan_turbulent_1962}.
We emphasize that the scaling is steeper than in the beginning of the ultimate regime of RB, where the logarithmic corrections from the turbulent thermal BLs yield an effective $\nusselt\sim\rayleigh^{0.38}$ scaling~\cite{lohse_ultimate_2024}. 
In contrast, 
in SID thermal BLs are eliminated by the reservoirs, and the entire channel is analogous to homogeneous RB flow~\cite{schmidt_axially_2012}.

We now turn our attention to the hydraulic dimensionless mass flux
\begin{equation}
    Q_m \equiv \frac{2\bracket{u_x(\rho-\rho_0)}}{U_0 \, \Delta \rho} = \frac{2(\nusselt-1)}{(\rayleigh \, \prandtl)^{1/2}}, 
\end{equation}
a metric widely used in the hydraulics and stratified flow communities. 
According to classical hydraulic control theory for two-layer steady exchange flows, $Q_m\rightarrow 1/2$  in the inviscid limit  $\reynolds,\rayleigh\rightarrow\infty$~\cite{stommel_control_1953,wood_lock_1970,armi_hydraulics_1986,dalziel_two-layer_1991}. 
\Cref{fig:qm} reveals a non-monotonic dependency of $Q_m$ on Ra.
In the low-Ra regime, $Q_m$ approaches its maximal value of about $0.5$ where the turbulence- or wave-driven exchange of
the  scalar across the density interface is minimal.
As Ra increases, the flow becomes more turbulent and mixed, and $Q_m$ decreases.
It eventually plateaus in the ultimate regime towards a constant well below 0.5. 
Similar behavior has been documented in SID experiments with salt ($\prandtl_s\approx700$), where our and their Re are comparable, but their $\rayleigh\approx 10^9$ -- $10^{11}$ are two orders of magnitude higher than ours~\cite{lefauve_buoyancy-driven_2020}. 
In those experiments, $Q_m$ converges to a larger asymptotic value close to 0.5 (see their fig. 6). 
Our data suggest a slight logarithmic uptrend in the ultimate regime, possibly hinting at $Q_m \rightarrow 0.5$ as $\rayleigh\rightarrow\infty$. 
On the other hand, the $\mathrm{Nu}\sim\mathrm{Ra}^{1/2}$ scaling is an  upper bound for turbulent convection at finite Pr~\cite{choffrut_upper_2016,lohse_ultimate_2024}, implying that $Q_m$ is independent of Ra in the ultimate regime. 
Further investigation into the asymptotic behavior at large Ra and Re, as well as the influence of Pr would be valuable to better understand the (dis)similarities between SID flow and the broader class of wall-bounded turbulent convection.

\subsection{\label{sec:mixing}Mixing and energy budget}
\begin{figure*}[htbp]
    \centering
    \subfloat{\includegraphics[height=2.1in]{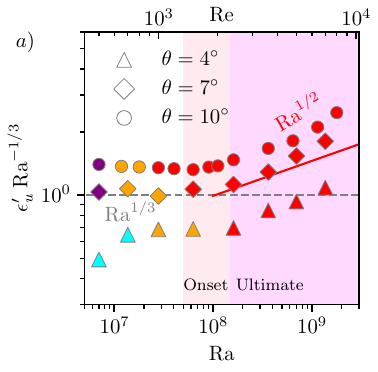}\label{fig:epsilon}}
    \subfloat{\includegraphics[height=2.1in]{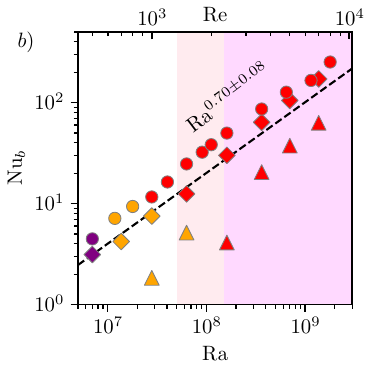}\label{fig:nub}}
    \subfloat{\includegraphics[height=2.1in]{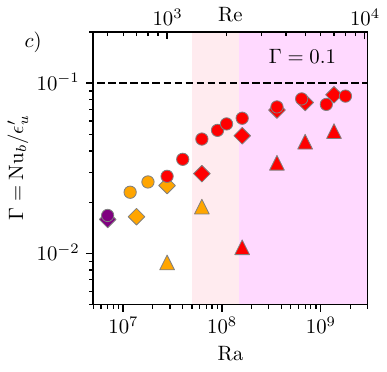}\label{fig:gamma}}
    \caption{(a) Log-log plot of the dimensionless dissipation $\epsilon_u' = \langle \epsilon _u\rangle \frac{H^4\prandtl^2}{\nu^3\rayleigh}$ as a function of Ra, compensated by $\rayleigh^{1/3}$. The dashed line guides the scaling of $\rayleigh^{1/3}$ and the red solid line the ultimate $\rayleigh^{1/2}$. (b) The buoyancy flux $\nusselt_b$ plotted against Ra. The dashed line guides the measured slope of $\nusselt_b\sim \rayleigh^{0.70}$. (c) The mixing efficiency $\Gamma = \nusselt_b/\epsilon_u'$. The dashed line suggests the asymptotic value of $\Gamma=0.1$. The first two laminar data points for $\theta = 4^\circ$ are excluded from (b) and (c). }
\end{figure*}
In contrast to RB convection, where the relation between $\nusselt$ and the kinetic energy dissipation $\bracket{\epsilon_u}\equiv\nu\langle|\mathbf{\nabla u}|^2\rangle$ is exact and closed~\cite{lohse_ultimate_2024}, such a relation exists neither in SID nor in vertical convection~\cite{ng_vertical_2015}. The energy budget equation is
\begin{equation} \label{eqn:epsilon}
    \begin{split}
        \langle \epsilon _u\rangle - \Phi_u= \frac{\nu^3}{H^4} \frac{\rayleigh}{\prandtl^2} \big[\sin\theta &\: (\nusselt-1) -   \cos\theta\: (\nusselt_b-1) \big],
    \end{split}
\end{equation}
where 
\begin{equation}
    \nusselt_b \equiv \frac{\bracket{u_z(\rho-\rho_0)}}{\kappa \Delta\rho /H} + 1
\end{equation}
 is the buoyancy flux quantifying vertical mixing in the duct's frame of reference and $\Phi_u = (1/V)\langle u_x |\bm{u}|^2/2 + u_x p \rangle_A\big|_{R-L}$ is the net convective momentum flux ($\langle \cdot \rangle_A$ denoting averaging over the duct cross-section of $H^2$ and time, and $\cdot|_{R-L}$ is the difference between the right and left duct ends).
The dimensionless dissipation 
\begin{equation}
    \epsilon_u'\equiv \langle \epsilon _u\rangle \frac{H^4\prandtl^2}{\nu^3\rayleigh},
\end{equation}
shown in \cref{fig:epsilon}, exhibits a steeper $\rayleigh$ scaling in the ultimate regime, with $\epsilon_u'\sim\rayleigh^{1/3}$ before and $\epsilon_u'\sim\rayleigh^{1/2}$ after the transition, reflecting
 the $\nusselt(\rayleigh)$ scalings.
The edge momentum flux $|\Phi_u| \ll \bracket{\epsilon_u}$ is very small according to our numerical results and can thus be neglected in \cref{eqn:epsilon} (see \cref{sec:appendix}, \cref{tab:2}). 
The buoyancy flux $\nusselt_b$, shown in \cref{fig:nub}, exhibits an effective scaling $\nusselt_b \sim \rayleigh^{0.70}$. 
In contrast to RB convection, where $\nusselt_b = 0$ due to homogeneity along the direction normal to the lateral walls, SID stratification leads to a finite buoyancy flux.
The negative sign in front of $\nusselt_b$ in \cref{eqn:epsilon} indicates that the kinetic energy generated by the buoyancy $\nusselt$ is consumed by both dissipation $\epsilon_u$ and mixing  $\nusselt_b$ in the stratified layer.
Notably, no clear transition is observed in the $\nusselt_b(\rayleigh)$ scaling, consistent with our conjecture that the ultimate transition in SID is associated with the top and bottom wall kinetic BLs (c.f. \cref{sec:BL}), where $|\partial_z\rho|\approx 0$ and $|u_z|\ll 1$, contributing negligibly to $\nusselt_b$.

In \cref{fig:gamma}, we plot the ratio 
\begin{equation}
    \Gamma \equiv \frac{\nusselt_b}{\epsilon_u'},
\end{equation}
commonly referred to as the flux parameter, a key quantity in the ocean mixing literature \cite{gregg_mixing_2018} that characterizes mixing efficiency—essentially quantifying the `taxation' of stratification on turbulent dissipation~\cite{caulfield_open_2020}. 
We observe that $\Gamma$ increases from $\approx 0.01$ and appears to asymptote to $\approx 0.1$ at large Ra for all three angles $\theta$. 
The value aligns with previous stratified turbulence DNS in a triply periodic domain~\cite{feraco_vertical_2018} and with previous  SID experiments~\cite{lefauve_experimental_2022-1} and  DNS \cite{zhu_stratified_2023}.
Notably, it is significantly lower than the widely adopted $\Gamma \approx 0.2$ in the oceanography literature~\cite{osborn_estimates_1980,gregg_mixing_2018,caulfield_open_2020} for reasons that remain to be understood.

\subsection{\label{sec:CVC}Analogy to convection in a vertical channel (CVC)}
\begin{figure*}[htbp]
    \centering
    \subfloat{\includegraphics[width=2.1in]{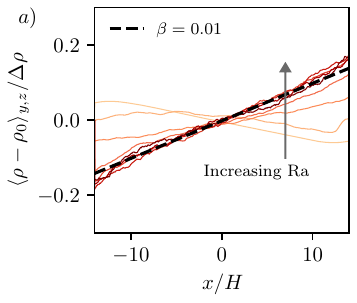}\label{fig:rhox}}
    \subfloat{\includegraphics[width=2.3in]{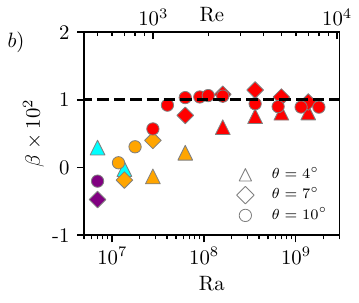}\label{fig:beta}}
    
    \subfloat{\includegraphics[width=4in]{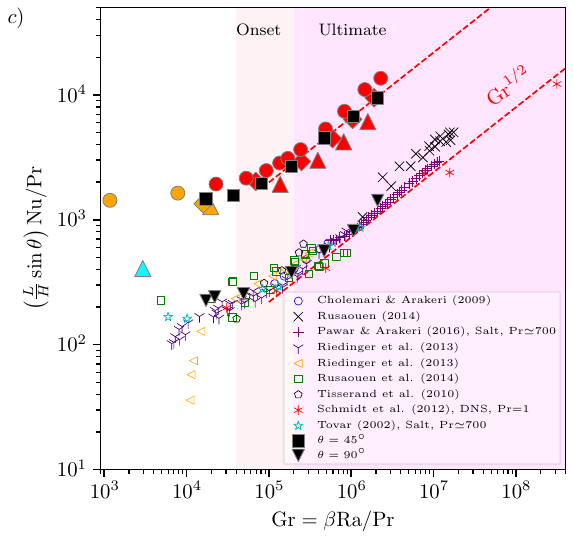}\label{fig:large_inclination}}
    \caption{
    (a) Mean streamwise density profiles in $x/H\in[-14,14]$ of $\theta=7^\circ$, satisfying $(\rho|_{x \approx \pm 20H} - \rho_0) / \Delta\rho = \pm 0.5$ in the reservoirs. Deeper shade of color indicates higher Ra. (b) Density gradient $\beta$ for all three angles. The dashed lines in (a,b) marks the asymptotic value of $\beta=0.01$ at large Ra.  (c) Current SID DNS results (large colored markers, showing only data with positive $\beta$) compared with previous CVC data from the literature ($\theta=90^\circ$)~\cite{Tovar2002,cholemari_axially_2009,rusaouen_echanges_2014,rusaouen_laminar_2014,riedinger_heat_2013,tisserand_convection_2010,pawar_two_2016,schmidt_axially_2012} and our additional results of large inclinations (black markers). Previous data are from experiments with heat in water ($\prandtl \approx 7$), unless otherwise indicated in the legend. The vertical axis is compensated by $(L/H)\sin\theta \:\prandtl^{-1}$.}
\end{figure*}
We now seek to understand the origin of the observed ultimate scaling $\nusselt\sim \rayleigh^{1/2}$ through the analogy to CVC, where this scaling is widely reported due to a linear density gradient driving the turbulence~\cite{riedinger_heat_2013,pawar_two_2016,castaing_turbulent_2017}. We plot the streamwise dimensionless density (or buoyancy) profiles $\frac{\langle \rho - \rho_0 \rangle_{y,z,t}}{\Delta\rho}(x)$ for $\theta=7^\circ$ in \cref{fig:rhox}, where $\langle \cdot\rangle_{y,z,t}$ denotes averaging over $y$, $z$, and time.
As $\rayleigh$ increases, these profiles become linear in $x$ with a slope of approximately 0.01. 
In \cref{fig:beta}, we show the measured dimensionless mean density gradient 
\begin{equation}
    \beta \equiv H\frac{\di}{\di x} \left(\frac{\langle \rho - \rho_0 \rangle_{y,z,t}}{\Delta\rho}\right)
\end{equation}
for all three angles $\theta$, where the asymptotic value of $\beta$ approaches 0.01 regardless of $\theta$.
The interpretation is that, as mixing develops within the duct, turbulent convection becomes dominant in the ultimate regime, sustained by the streamwise linear density gradient, resulting in the $\nusselt\sim \rayleigh^{1/2}$ scaling consistent with CVC.
At lower Ra where turbulence is not fully developed, we observe that some cases exhibit negative density gradients ($\beta < 0$).
These flows are hydraulically subcritical, as explained by hydraulic control theory~\cite{lefauve_regime_2019}, primarily governed by interfacial waves and linear instabilities (as visualized in \cref{fig:flow}b-d), rather than by turbulent convection.

To quantify the effective buoyancy driving strength in the duct, we define a Grashof number based on the mean density gradient $\beta$
\begin{equation}
    \mathrm{Gr} \equiv \beta\frac{g(\Delta\rho/\rho_0) H^3}{\nu^{2}} = \beta \frac{\rayleigh}{\prandtl}.
\end{equation}
In~\cref{fig:large_inclination}, we plot our DNS results alongside previous experimental~\cite{Tovar2002,cholemari_axially_2009,rusaouen_echanges_2014,rusaouen_laminar_2014,riedinger_heat_2013,tisserand_convection_2010,pawar_two_2016} and numerical~\cite{schmidt_axially_2012} studies of CVC, where Nu is compensated by $\prandtl^{-1}$ and $(L/H)\sin\theta$ (the effective nondimensional transport distance along gravity).
All data follow a similar $\mathrm{Gr}^{1/2}$ scaling towards the ultimate regimes, tho the effective scalar transport decreases significantly for $\theta=90^\circ$.
SID exhibits higher scaling prefactors possibly due to the mean velocity gradient aligning with the stratification, approximating a two-layer configuration and thereby enhancing the transport.
Interestingly, the onset $\mathrm{Gr}\approx10^5$ appears to be universal for both SID (small $\theta$) and CVC ($\theta=90^\circ$). 
We further present two additional sets of simulations at \(\theta=45^\circ\) and \(90^\circ\) shown as black markers in \cref{fig:large_inclination}. For these large inclinations, the stratification is weak or absent and the system approaches CVC. 
Notice that in~\cite{Tovar2002,cholemari_axially_2009,rusaouen_echanges_2014,rusaouen_laminar_2014,riedinger_heat_2013,tisserand_convection_2010,pawar_two_2016,schmidt_axially_2012}, the length scale in Nu is the duct length $L$. We use the duct height $H$ instead in this work. Our $\theta=90^\circ$ data collapse nicely with the CVC data after being compensated by the duct aspect ratio $L/H=30$. 

Despite similarities in highly turbulent regimes, the routes to turbulence in the two systems are fundamentally different.
CVC transitions through a series of exponentially growing unstable elevator modes~\cite{calzavarini_exponentially_2006,schmidt_axially_2012}. 
In contrast, the transition of SID flow is initially associated with supercritical bifurcations by a family of linear instabilities of the stratified shear layer~\cite{atoufi_stratified_2023,zhu_long-wave_2024}.
Lefauve~\cite{lefauve_geophysical_2024} further hypothesized the existence of a subcritical route from intermittency to fully developed turbulence, similar to other initially supercritical systems like RB and Taylor-Couette flow.
In the remainder of this paper, we present evidence supporting this subcritical route to ultimate turbulence mediated by turbulence developing within the kinetic BLs.

\begin{figure}[ht]
    \includegraphics[width=3.2in]{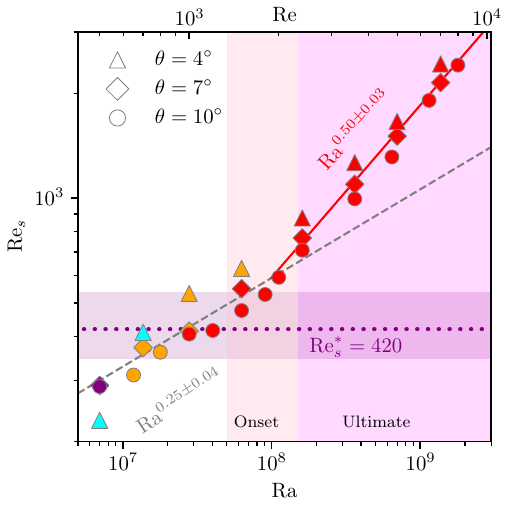}
    \caption{\label{fig:res} Log-log plot of shear Reynolds number $\reynolds_s$ against $\rayleigh$. The purple horizontal dotted line and strip mark  $\reynolds_s^*=420$ around which the laminar BLs are expected to become turbulent, emphasizing the subcritical nature of the transition, which occurs over a range of $\reynolds_s$. The fitted slopes match the expected  $\reynolds_s\sim\rayleigh^{1/4}$ (dashed) for laminar BL and $\reynolds_s\sim\rayleigh^{1/2}$ (red solid) for turbulent BL.}
\end{figure}

\section{Turbulent boundary layers and transition\label{sec:BL}}
\subsection{\label{sec:reynolds}Shear Reynolds number}
The key parameter to characterize the transition of kinetic BL structure is the shear Reynolds number
\begin{equation}
    \reynolds_s \equiv \frac{\max\langle u_x\rangle_{x,y,t} \delta^*}{\nu},
\end{equation}
based on the displacement BL thickness $\delta^*  = \int_0^{z_\mathrm{max}} \left( 1- \frac{\bracket{u_x}_{x,y,t} (z)}{\max\langle u_x\rangle_{x,y,t}}\right)\di z$~\cite{landau_fluid_1987}, where $z_\mathrm{max}$ corresponds to the location of $\max\langle u_x\rangle_{x,y,t} $, associated with the BLs on top and bottom walls.
In \cref{fig:res}, we plot $\reynolds_s$ as a function of $\rayleigh$.
In the low-Ra regime, we find that the measured slope $\reynolds_s \sim \rayleigh^{0.25} \sim \reynolds^{0.5}$, matches with the expected scaling based on the laminar BL thickness $\delta^*/H\sim \rayleigh^{-1/4}\sim \reynolds^{-1/2}$. 
In the ultimate regime, the scaling shifts to $\reynolds_s \sim \rayleigh^{1/2} \sim \reynolds$ for turbulent BLs. 
Landau and Lifshitz~\cite{landau_fluid_1987} reported the estimate of $\reynolds_s^* \approx  420$ for the range of the transition to a turbulent BL (a detailed discussion can be found in~\cite{lohse_ultimate_2024}).
We observe that indeed $\reynolds_s \gtrsim 420$ coincides with the onset region (highlighted in orange) where the $\nusselt$ scaling changes (\cref{fig:nux,fig:nux_comp}).
This suggests that the transition to ultimate turbulence may be caused by changes in the BLs, although a coincidence cannot be ruled out.

\begin{figure*}[ht]
    \includegraphics[width=\textwidth]{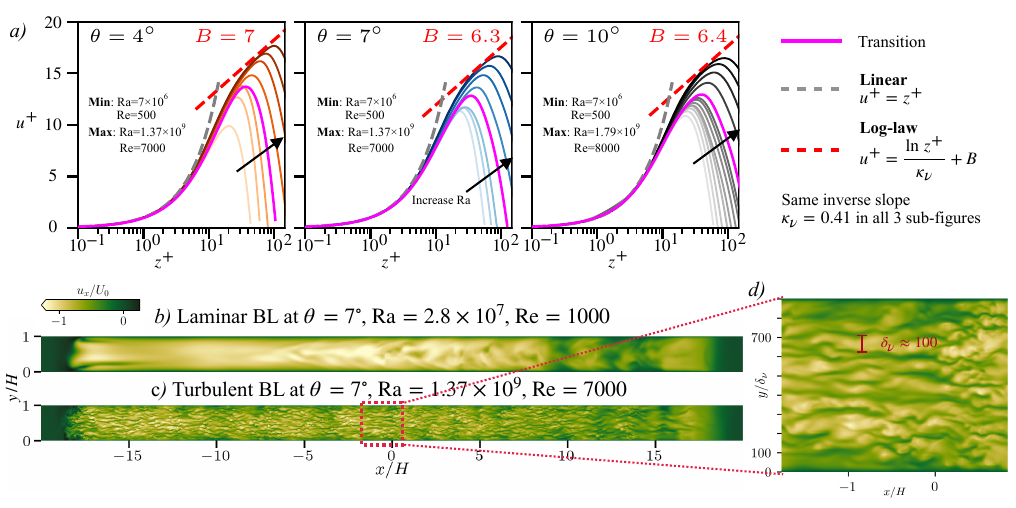}
    \caption{\label{fig:bl}(a) Overall streamwise velocity $\bracket{u_x|_{y=H/2}}_{x,t}(z)$ profiles in wall-normal units. Deeper colors represent higher $\rayleigh$. The dashed lines correspond to the canonical 3D wall-bounded turbulence profile: (grey) the linear viscous sublayer, and (red) the log-law with inverse slope $\kappa_\nu = 0.41$ and $B$ varies slightly for different $\theta$. (b) Flow visualization (bottom view) of streamwise velocity $u_x$ at $z\approx 12\delta_\nu$ inside the BL before the transition, and (c) after the transition, zoomed in (d). The $y$-axes are scaled up by 2 in (b-d).}
\end{figure*}

\subsection{\label{sec:profiles}Streamwise velocity profiles}

A hallmark feature of a turbulent BL is the log-law of the wall. We plot the averaged streamwise velocity profiles near $z=0$ (bottom wall) at $y=H/2$ (spanwise mid-plane) in \cref{fig:bl}a. 
The profiles are shown in the wall-normal units with $u^+ \equiv \bracket{u_x}_{x,t}/u_\tau$, where $u_\tau = \sqrt{\nu \partial_z \bracket{u_x}_{x,t}|_{z=0}}$ is the friction velocity, and
the spatial coordinates $z^+ \equiv z/\delta_\nu$, with $\delta_\nu = \nu/u_\tau$ being the viscous unit.
The magenta curves mark the profiles near the ultimate transition where the BLs have just turned turbulent. 
At large $\rayleigh$, we notice that the curves converge to the shape of canonical 3D wall turbulence, comprising the viscous linear sublayer and by the short logarithmic region $u^+ = (1/\kappa_\nu) \ln z^+ + B$ with $\kappa_\nu=0.41$ agreeing with the von K\'{a}rm\'{a}n constant. The parameter $B$ differs slightly for different $\theta$.
We note that the logarithmic regions are not fully developed even at our highest $\rayleigh$, spanning only about half a decade in $z^+$. 
This is not surprising: in SID, the logarithmic region is immediately followed by a negative mean velocity gradient in the bulk, arising from the counter-directional nature of the exchange flows, rather than by the free-streaming wake region of canonical turbulent boundary layers. This likely suppresses the log-law from extending further into $z^+$.

To further appreciate the features in the BLs in different regimes, we plot snapshots of the streamwise velocity $u_x$ near the buffer layer $z^+\approx 12$. 
The laminar BL in \cref{fig:bl}b exhibits an overall smooth profile, with disturbances being generated upstream and dissipated downstream. The turbulent BL in \cref{fig:bl}c is qualitatively different, exhibiting meandering streaks across the entire duct. 
These streaks are another hallmark of turbulent BLs, well-documented in canonical turbulent BL literature~\cite{smits_highreynolds_2011}. 
In the zoom-in \cref{fig:bl}d, we observe that the spanwise spacings of these streaks are around 100-200 viscous units, consistent with previous studies on wall turbulence~\cite{kline_structure_1967,smits_highreynolds_2011}.

\begin{figure*}[htbp]
   \includegraphics[width=\textwidth]{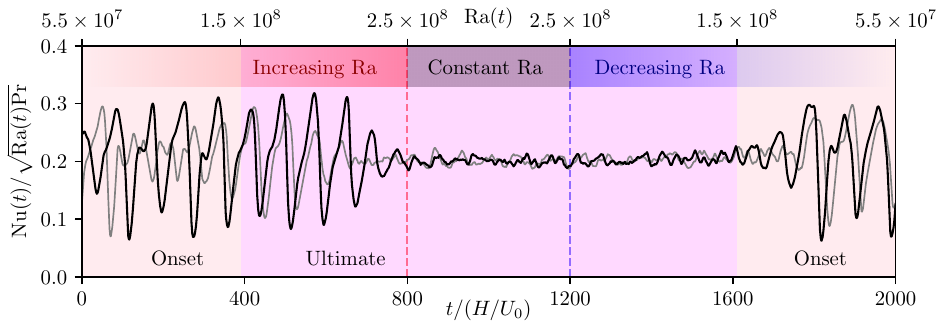}
    \caption{\label{fig:hyster}
    Hysteresis of the instantaneous Nu (compensated by
$\sqrt{\rayleigh\prandtl}$) plotted against dimensionless time.
The upper axis shows the imposed $\rayleigh(t)$: a slow linear ramp up, a constant hold, and an equally slow ramp down.
Two independent realizations at $\theta=4^\circ$ are shown (black and grey). The up- and down-ramps span the transition Ra-range towards the ultimate SID regime, and the regime colors match \cref{fig:nux_comp}}
\end{figure*}
\subsection{\label{sec:transition}Nature of the transition to ultimate SID}
The non-normal-nonlinear nature of the transition is demonstrated in figure \ref{fig:hyster}. 
We run two time-dependent realizations spanning the onset and ultimate range of $\mathrm{Ra}\approx 5.5\times10^7$ -- $2.5\times10^8$ ($\mathrm{Re}\approx 1400$ -- $3200$; see \cref{fig:nux_comp}). 
Starting in the intermittent regime, the compensated transport $\mathrm{Nu}(t)/\sqrt{\mathrm{Ra}(t)\mathrm{Pr}}$ shows strong intermittency, with peaks during the quiet phase (\cref{fig:flow}c) and valleys during the active phase (\cref{fig:flow}d). 
As Ra is slowly increased towards $\approx2.5\times10^8$ (black curve), the fluctuations decrease significantly, marking the entry into the ultimate regime.
During the ramp down, however, the ultimate regime persists until a much lower $\mathrm{Ra}\approx1.5\times10^8$, after which intermittency reappears.
The second realization (grey) transitions at different Ra, and the up- and down-ramp thresholds also differ from each other.
This realization-dependence and hysteresis reflect that there is no single sharp Ra for the ultimate regime, and thus that the transition is subcritical and hysteretic.

Linear stability analysis suggests SID flow becomes linearly unstable for $\reynolds\lesssim 650$, $\rayleigh\lesssim 1.2\times 10^6$~\cite{zhu_long-wave_2024}, well below the observed turbulent BL and ultimate transition.
This finding suggests that the nonlinear subcritical shear instability developed within the BLs is crucial to trigger ultimate turbulent convection in SID. 
This contrasts with homogeneous RB~\cite{schmidt_axially_2012}, where no evidence of turbulent BLs on the lateral walls has been reported.

\section{\label{sec:conclusion}Conclusions and outlook}
In this work, we presented the first 3D DNS of SID flow in the ultimate regime. The transition at $\rayleigh\approx 10^8$ is marked by an  enhancement in mass transfer efficiency to the asymptotic ultimate scaling of $\nusselt\sim\rayleigh^{1/2}$. 
We interpret this as the point where turbulent convection becomes the dominant driving mechanism, sustained by the formation of a streamwise uniform density gradient, analogous to the behavior observed in homogeneous RB.
Simultaneously, the shear Reynolds number at the top and bottom walls exceeds $\reynolds^*_s\approx 420$, triggering turbulent BLs through a non-normal-nonlinear subcritical route. 
The Ra-range for the onset of the ultimate regime coincides with the turbulent BL transition, independently of the duct inclination $\theta$, despite clear variations in global flow regimes with $\theta$, supporting the interpretation that the transition is associated with the BLs.
From a broader fluid dynamics perspective, SID flow offers a distinct internal wall shear flow configuration characterized by two extrema in the mean velocity profile, contrasting with canonical wall turbulence which exhibit either no extremum (plane Couette) or a single  extremum (Poiseuille).

Many questions remain open. 
Although the asymptotic $Q_m = 1/2$ predicted by inviscid two-layer hydraulics coincides with the scaling $\nusselt\sim\rayleigh^{1/2}$, our results show that $Q_m < 0.5$ in the current range of $\rayleigh\approx 10^7$ -- $10^9$. 
It remains unclear whether $Q_m$ will remain below 0.5 indefinitely or eventually approach 0.5 as $\rayleigh\rightarrow\infty$. 
Furthermore, our current interpretation is unlikely to hold for very small $\theta$ or in a horizontal duct ($\theta \approx 0^\circ$), where the streamwise convective driving is weak or absent compared to horizontal hydrostatic pressure, even in the turbulent regime. Whether horizontal SID flow can exhibit the ultimate regime remains to be investigated.

Our results highlight universalities among various wall-bounded convective flows. Beyond fundamental fluid dynamics, our scaling laws for turbulent transport and energetics, including mixing efficiency, may help inform high-Re and high-Ra closure models in geophysics and engineering.

\begin{acknowledgments}
We thank Dr. C. Howland, G. Vacca and Dr. K. Zhong for fruitful discussions on numerical techniques and boundary layers
and Prof. J. Salort for sharing data of convection in a vertical channel. AL was supported by a NERC Independent Research Fellowship (NE/W008971/1). 
We acknowledge the EuroHPC Joint Undertaking for awarding the grant EHPCREG-2023R03-178 CPU hours on the EuroHPC supercomputer Discoverer at the Sofia Tech Park in Bulgaria; the project `2024.056-Convection and phase-change in turbulent geophysical flows' of the research program ``Computing Time on National Computing Facilities" co-funded by the Dutch Research Council (NWO); and the Gauss Center for Supercomputing e.V. for project ID pr74sa on the GCS Supercomputer SuperMUC-NG at the Leibniz Supercomputing Center. 
AL and DL also acknowledge the NSF- and ONR-funded Geophysical Fluid Dynamics Program at the Woods Hole Oceanographic Institution, where this work was initiated.
\end{acknowledgments}

\bibliography{apssamp}

\newpage
\appendix
\section{Numerical details and data\label{sec:appendix}}
We solved the following nondimensionalized equations which are the basis for the numerical simulations:
\begin{subequations}
\label{eqn:NS_numerical}
    \begin{align}
        \bm{\nabla}\cdot\bm{u} &= 0 \\
        \partial_t \bm{u} + (\bm{u}\cdot \bm{\nabla})\bm{u} &= - \bm{\nabla} p  + \sqrt{\frac{\prandtl}{\rayleigh}}\bm{\nabla}^2\bm{u} + \rho \bm{\hat{e}}_\theta - \bm{F}_u\\
        \partial_t \rho + (\bm{u}\cdot \bm{\nabla})\rho &= \sqrt{\frac{1}{\prandtl\rayleigh}} \bm{\nabla}^2 \rho- F_\rho.
    \end{align}
\end{subequations}
Following \cite{zhu_stratified_2023}, the artificial restoring forces $\bm{F}_u$ and $F_\rho$ are given by
\begin{align}
\bm{F}_u &\equiv F_u \bm{u} \equiv \left[ 
  \frac{1 - \tanh\left(\frac{8}{l_f}\left(x + \frac{L - l_f}{2}\right)\right)}{\eta_u} 
  + \frac{1 + \tanh\left(\frac{8}{l_f}\left(x - \frac{L - l_f}{2}\right)\right)}{\eta_u} 
\right] \bm{u}, \\
F_\rho &\equiv
  \frac{1 - \tanh\left(\frac{8}{l_f}\left(x + \frac{L - l_f}{2}\right)\right)}{\eta_\rho} \left(\rho - \frac{1}{2}\right) 
  + \frac{1 + \tanh\left(\frac{8}{l_f}\left(x - \frac{L - l_f}{2}\right)\right)}{\eta_\rho} \left(\rho + \frac{1}{2}\right).
\end{align}
$l_f$ is the streamwise length of influence of the forcing and $L$ is the simulation domain length in $x$. The parameters $\eta_u$ and $\eta_\rho$ control the restoring force strength. The parameters used for each simulation are provided in \cref{tab:1}.

\begin{figure}[htbp]
    \centering
    \includegraphics[width=0.6\linewidth]{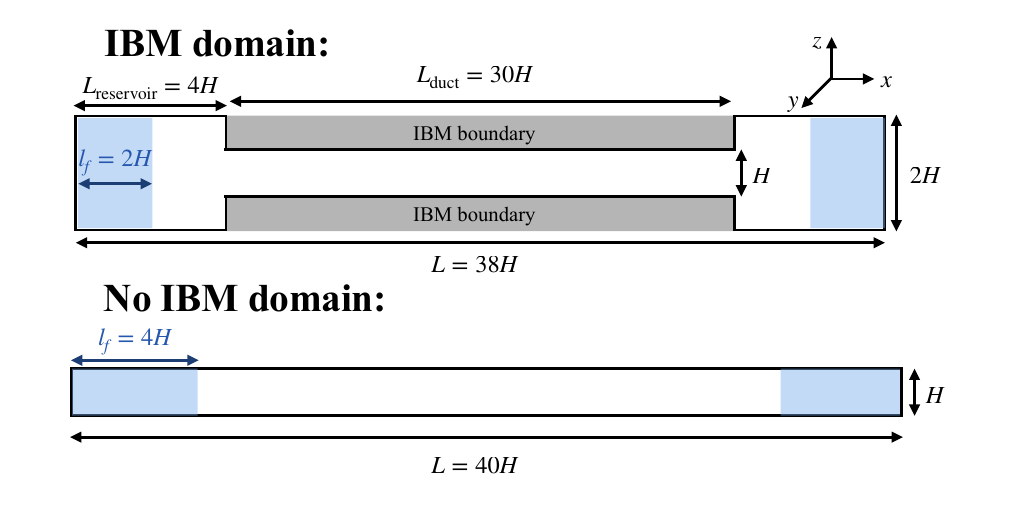}
    \caption{Sketch of simulation domain geometry. Both geometries have thickness $H$ in $y$ direction. Please note the figures are not on scale.}
    \label{fig:1}
\end{figure}

The detailed simulation domain geometry is illustrated in \cref{fig:1}. The $z$ wall-clustered grid scheme for velocity fields in \cref{tab:1} follows~\cite{ceci_natural_2023}. The time-averaged results are available in \cref{tab:2}. The two data points shown in the main text: $\theta=10^\circ, \rayleigh=1.61\times 10^8$ and $\theta=10^\circ, \rayleigh=3.63\times 10^8$ are the averages of IBM and non-IBM geometry simulation results

\begin{turnpage}
\begin{table}
    \centering
    \resizebox{1.2\textwidth}{!}{\begin{tabular}{c|c|c|c|c|c|c|c|c|c|c|c|c}
        $\theta$ & $\rayleigh$ & $\reynolds$ & IBM & $N_z\times N_x \times N_y $ & $N_z^r\times N_x^r \times N_y^r $ (Uniform) & $N_{BL}$ & $\Delta z^+_{1}$ & $L$ & $l_f$ & $\eta_u$ & $\eta_\rho$ & Regime\\
        \hline\hline
        $10^\circ$ & $7\times10^6$ & 500 & Yes & $256\times2048\times 128$ (Uniform)& $256\times2048\times 128$ & 23 & 0.42 & 38&2 &10 &0.2 & W\\
        $10^\circ$ & $1.18\times10^7$ & 650 & Yes & $256\times2048\times 128$ (Uniform)& $256\times2048\times 128$ & 20 & 0.50& 38&2&10&0.2 & I\\
        $10^\circ$ & $1.79\times10^7$ & 800 & Yes & $256\times2048\times 128$ (Uniform)& $256\times2048\times 128$ & 19 & 0.59& 38&2&10&0.2 & I\\
        $10^\circ$ & $2.8\times10^7$ & 1000 & Yes & $256\times2048\times 128$ (Uniform)& $256\times2048\times 128$ & 17 & 0.70& 38&2&10&0.2 & T\\
        $10^\circ$ & $4.03\times10^7$ & 1200 & Yes & $256\times3072\times 128$ (Uniform)& $256\times3072\times 128$ & 15 & 0.81& 38&2&10&0.2 & T\\
        $10^\circ$ & $6.3\times10^7$ & 1500 & Yes & $256\times3072\times 128$ (Uniform)& $256\times3072\times 128$ & 14 & 0.96& 38&2&10&0.2 & T\\
        $10^\circ$ & $9.07\times10^7$ & 1800 & Yes & $384\times3072\times 128$ (Uniform)& $384\times3072\times 128$ & 19 & 0.75& 38&2&10&0.2 & T\\
        $10^\circ$ & $1.12\times10^8$ & 2000 & Yes & $384\times3072\times 128$ (Uniform)& $384\times3072\times 128$ & 19 & 0.81& 38&2&10&0.2 & T\\
        $10^\circ$ & $1.61\times10^8$ & 2400 & Yes & $512\times4096\times 192$ (Uniform)& $512\times4096\times 192$ & 24 & 0.70& 38&2&10&0.2 & T\\
        $10^\circ$ & $3.63\times10^8$ & 3600 & Yes & $768\times6144\times 256$ (Uniform)& $768\times6144\times 256$ & 32 & 0.66& 38&2&10&0.2 & T\\
        $10^\circ$ & $1.61\times10^8$ & 2400 & No & $256\times4096\times 256$ ($z$ wall-clustered)& $256\times4096\times 256$ & 41 & $ 5.03 \times 10^{-3}$& 40&4&5&0.5 & T\\
        $10^\circ$ & $3.63\times10^8$ & 3600 & No & $512\times6144\times 384$ ($z$ wall-clustered)& $512\times6144\times 384$ & 71 & $ 2.36 \times 10^{-3}$& 40&4&5&0.5 & T\\
        $10^\circ$ & $6.45\times10^8$ & 4800 & No & $512\times6144\times 384$ ($z$ wall-clustered)& $512\times6144\times 384$ & 68 & $ 2.75 \times 10^{-3}$& 40&4&5&0.5 & T\\
        $10^\circ$ & $1.15\times10^9$ & 6400 & No & $768\times8192\times 512$ ($z$ wall-clustered)& $768\times8192\times 512$ & 100 & $ 1.91 \times 10^{-3}$& 40&4&5&0.5 & T\\
        $10^\circ$ & $1.79\times10^9$ & 8000 & No & $768\times8192\times 512$ ($z$ wall-clustered)& $768\times8192\times 512$ & 99 & $ 2.21 \times 10^{-3}$& 40&4&5&0.5 & T\\
        \hline
        $7^\circ$ & $7\times10^6$ & 500 &No  & $96\times1024\times 64$ ($z$ wall-clustered)& $192\times2048\times 128$ & 27 & $ 2.03 \times 10^{-2}$& 40&4&5&0.5 & W\\
        $7^\circ$ & $1.32\times10^7$ & 700 & No & $96\times1024\times 64$ ($z$ wall-clustered)& $192\times2048\times 128$ & 26 & $ 2.42 \times 10^{-2}$& 40&4&5&0.5 & I\\
        $7^\circ$ & $2.8\times10^7$ & 1000 & No & $128\times1536\times 96$ ($z$ wall-clustered)& $256\times3072\times 192$ & 30 & $ 1.84 \times 10^{-2}$& 40&4&5&0.5 & I\\
        $7^\circ$ & $6.3\times10^7$ & 1500 & No & $128\times1536\times 96$ ($z$ wall-clustered)& $256\times3072\times 192$ & 27 & $ 2.21 \times 10^{-2}$& 40&4&5&0.5 & T\\
        $7^\circ$ & $1.61\times10^8$ & 2400 & No & $192\times2048\times 128$ ($z$ wall-clustered)& $384\times4096\times 256$ & 34 & $ 1.44 \times 10^{-2}$& 40&4&5&0.5 & T\\
        $7^\circ$ & $3.63\times10^8$ & 3600 & No & $192\times2048\times 128$ ($z$ wall-clustered)& $384\times4096\times 256$ & 32 & $ 1.69 \times 10^{-2}$& 40&4&5&0.5 & T\\
        $7^\circ$ & $7\times10^8$ & 5000 & No & $256\times3072\times 192$ ($z$ wall-clustered)& $512\times6144\times 384$ & 39 & $ 1.33 \times 10^{-2}$& 40&4&5&0.5 & T\\
        $7^\circ$ & $1.37\times10^9$ & 7000 & No & $256\times3072\times 192$ ($z$ wall-clustered)& $512\times6144\times 384$ & 38 & $ 1.45 \times 10^{-2}$& 40&4&5&0.5 & T\\
        \hline
        $4^\circ$ & $7\times10^6$ & 500 & No & $96\times1024\times 64$ ($z$ wall-clustered)& $192\times2048\times 128$ & 29 & $ 2.03 \times 10^{-2}$& 40&4&5&0.5 & L\\
        $4^\circ$ & $1.32\times10^7$ & 700 & No & $96\times1024\times 64$ ($z$ wall-clustered)& $192\times2048\times 128$ & 29 & $ 2.42 \times 10^{-2}$& 40&4&5&0.5 & L\\
        $4^\circ$ & $2.8\times10^7$ & 1000 & No & $128\times1536\times 96$ ($z$ wall-clustered)& $256\times3072\times 192$ & 34 & $ 1.84 \times 10^{-2}$& 40&4&5&0.5 & I\\
        $4^\circ$ & $6.3\times10^7$ & 1500 & No & $128\times1536\times 96$ ($z$ wall-clustered)& $256\times3072\times 192$ & 31 & $ 2.21 \times 10^{-2}$& 40&4&5&0.5 & I\\
        $4^\circ$ & $1.61\times10^8$ & 2400 & No & $192\times2048\times 128$ ($z$ wall-clustered)& $384\times4096\times 256$ & 39 & $ 1.44 \times 10^{-2}$& 40&4&5&0.5 & T\\
        $4^\circ$ & $3.63\times10^8$ & 3600 & No & $192\times2048\times 128$ ($z$ wall-clustered)& $384\times4096\times 256$ & 36 & $ 1.69 \times 10^{-2}$& 40&4&5&0.5 & T\\
        $4^\circ$ & $7\times10^8$ & 5000 & No & $192\times2048\times 128$ ($z$ wall-clustered)& $384\times4096\times 256$ & 34 & $ 1.98 \times 10^{-2}$& 40&4&5&0.5 & T\\
        $4^\circ$ & $1.37\times10^9$ & 7000 & No & $256\times3072\times 192$ ($z$ wall-clustered)& $512\times6144\times 384$ & 43 & $ 1.45 \times 10^{-2}$& 40&4&5&0.5 & T\\
        \hline
    \end{tabular}}
    \caption{ Values of the control parameters and resolutions of the simulations. $N_z$ is the velocity field resolution in the vertical direction (normal to the duct), $N_x$ streamwise, and $N_y$ spanwise. $N^r$ is the refined resolution for scalar field $\rho$. $N_{BL}$ is number of grid points in the boundary layer. $\Delta z^+_{1}$ is the distance of first grid point from the wall in wall-normal units.}
    \label{tab:1}
\end{table}
\end{turnpage}

\begin{table}
    \centering
    \resizebox{\textwidth}{!}{\begin{tabular}{c|c|c|c|c|c|c|c|c|c|c}
        $\theta$ & $\rayleigh$ & $\reynolds$ & $\nusselt$ & $\sigma_\nusselt$ & $\frac{\langle \epsilon_u \rangle}{\nu^3H^{-4}}\frac{\mathrm{Pr}^2}{\mathrm{Ra}}$ & $\sigma_{\epsilon_u}$ & $\frac{\langle\Phi_u\rangle}{\nu^3H^{-4}}\frac{\mathrm{Pr}^2}{\mathrm{Ra}}$ & $\reynolds_L$ & $\reynolds_s$ & $\tau$\\
        \hline\hline
        $10^\circ$ & $7\times10^6$ & 500 &1617.31 & $\pm 23.03 $&267.85 & $\pm 5.02 $&-8.59&1014.62&287.57 & 200\\
        $10^\circ$ & $1.18\times10^7$ & 650 &1919.97& $\pm 127.42 $&312.27& $\pm 23.47 $&-14.09&1238.02&310.38& 200\\
        $10^\circ$ & $1.79\times10^7$ & 800 &2192.74 & $\pm 183.20 $&356.37 & $\pm 35.77 $&-15.17&1476.64&360.90& 200\\
        $10^\circ$ & $2.8\times10^7$ & 1000 & 2591.32 & $\pm 160.16 $&410.15& $\pm 30.65 $&-28.36&1882.23&406.79& 200\\
        $10^\circ$ & $4.03\times10^7$ & 1200 &2889.55 & $\pm 36.59 $&458.19& $\pm 8.39 $&-27.45&2224.87&416.60& 200\\
        $10^\circ$ & $6.3\times10^7$ & 1500 &3324.60 & $\pm 36.60 $&525.94& $\pm 7.52 $&-27.03&2711.28&475.46& 200\\
        $10^\circ$ & $9.07\times10^7$ & 1800 &3818.07 & $\pm 33.63 $&611.01& $\pm 7.34 $&-20.18&3242.53&528.55& 100\\
        $10^\circ$ & $1.12\times10^8$ & 2000 &4164.90 & $\pm 26.74 $&664.13& $\pm 5.58 $&-21.38&3613.27&592.15& 100\\
        $10^\circ$ & $1.61\times10^8$ & 2400 &4793.24 & $\pm 28.41 $&778.74& $\pm 5.75 $&-5.02&4346.93&689.07& 100\\
        $10^\circ$ & $3.63\times10^8$ & 3600 &6851.08 & $\pm 40.96 $&1137.68& $\pm 10.97 $&27.69&6696.61&955.65& 100\\
        $10^\circ$ & $1.61\times10^8$ & 2400 &4996.74 & $\pm117.42$ & 824.72 & $\pm34.27$&101.11&4360.58&727.45 & 200\\
        $10^\circ$ & $3.63\times10^8$ & 3600 &7469.28 & $\pm120.83$ & 1243.65 & $\pm29.72$&208.71&6737.19&1034.11& 101\\
        $10^\circ$ & $6.45\times10^8$ & 4800 &9914.83 & $\pm117.54$ & 1568.99 & $\pm97.51$&209.02&9129.29&1313.24& 50\\
        $10^\circ$ & $1.15\times10^9$ & 6400 &14802.87 & $\pm506.44$ & 2208.70 & $\pm185.06$&111.04&12883.95&1908.52& 25\\
        $10^\circ$ & $1.79\times10^9$ & 8000 &18214.52 & $\pm729.95$ & 3004.30 & $\pm185.77$&560.66&16624.96&2409.56& 25\\
        \hline
        $7^\circ$ & $7\times10^6$ & 500&1617.83 & $\pm82.56$ & 197.81 & $\pm9.66$ &9.64 &937.20 &289.61& 200\\
        $7^\circ$ & $1.32\times10^7$ & 700&2101.43 & $\pm367.52$ & 256.59 & $\pm65.02$ &12.60 &1269.85 &372.18& 200\\
        $7^\circ$ & $2.8\times10^7$ & 1000&2570.95 & $\pm463.44$ & 300.32 & $\pm90.33$ &8.61&1612.27 &414.71& 400\\
        $7^\circ$ & $6.3\times10^7$ & 1500&3874.74 & $\pm649.94$ & 424.29 & $\pm123.86$ &-12.06&2638.98 &549.23& 400\\
        $7^\circ$ & $1.61\times10^8$ & 2400&5599.55 & $\pm460.51$ & 611.77 & $\pm95.37$ &15.71 &4261.51 &768.07& 200\\
        $7^\circ$ & $3.63\times10^8$ & 3600&8531.84 & $\pm182.43$ & 918.37 & $\pm30.34$ &62.12&6626.91 &1095.74 & 139\\
        $7^\circ$ & $7\times10^8$ & 5000&12165.58 & $\pm610.55$ & 1363.96 & $\pm166.36$ &183.37 &9562.43 &1506.41 & 83\\
        $7^\circ$ & $1.37\times10^9$ & 7000&18127.54 & $\pm1279.37$ & 2007.37 & $\pm311.96$ &291.23&14089.94 &2146.70 & 77\\ 
        \hline
        $4^\circ$ & $7\times10^6$ & 500&1359.54 & $\pm1.65$ & 94.20 & $\pm0.19$ &-4.78 &656.78 &230.06 &200\\
        $4^\circ$ & $1.32\times10^7$ & 700&2344.97 & $\pm34.57$ & 154.70 & $\pm1.01$ &-8.69&1162.11 &411.52&200\\ 
        $4^\circ$ & $2.8\times10^7$ & 1000 &3188.56 & $\pm312.16$ & 208.25 & $\pm28.44$ &-8.85&1734.70 &532.40&400\\
        $4^\circ$ & $6.3\times10^7$ & 1500&4248.12 & $\pm659.02$ & 274.13 & $\pm78.71$ &-7.30&2458.04 &628.35 &400\\
        $4^\circ$ & $1.61\times10^8$ & 2400 &6420.84 & $\pm816.47$ & 379.25 & $\pm91.40$ &-56.75&4041.98 &877.18&100\\
        $4^\circ$ & $3.63\times10^8$ & 3600&10013.86 & $\pm241.19$ & 602.31 & $\pm42.37$ &-36.95&6331.28 &1264.48&100 \\
        $4^\circ$ & $7\times10^8$ & 5000 & 14046.73 & $\pm93.29$ & 827.48 & $\pm23.89$ &-43.99 &8994.81 &1661.16&100\\
        $4^\circ$ & $1.37\times10^9$ & 7000&20416.35 & $\pm219.84$ & 1206.43 & $\pm39.36$ &-35.41 &12996.63 &2428.80&100\\
    \end{tabular}}
    \caption{Results from the numerical simulations. The standard deviations $\sigma$ are computed based on fluctuations over time. $\Phi_u$ is computed based on \cref{eqn:epsilon}, which is small compare to $\bracket{\epsilon_u}$, roughly comparable to the standard deviation $\sigma_{\epsilon_u}$. $\reynolds_L = \max\langle u_x\rangle_{x,y,t}H/\nu$ is the bulk Reynolds number. $\tau$ is the nondimensional time to get the statistics, scaled by $T_0 = H/U_0 = \sqrt{H/g(\Delta\rho/\rho_0)}$.}
    \label{tab:2}
\end{table}

\end{document}